\documentclass[a4paper,11pt]{article}
\usepackage{pos}
\usepackage{placeins}
\usepackage{booktabs}
\usepackage{float}
\usepackage{hyperref}
\usepackage[format=plain,labelfont=bf,textfont=it]{caption}
\title{Off-shell $t\bar{t}W^+j$ production at NLO QCD accuracy}

\author*[a]{Minos Reinartz\,}

\affiliation[a]{Institute for Theoretical Particle Physics and Cosmology, RWTH Aachen University\\
  Sommerfeldstr. 16, Aachen, Germany}

\emailAdd{minos.reinartz@rwth-aachen.de}

\abstract{We report on our recent full off-shell $pp \rightarrow t\bar{t}W^+ j+X$ calculation at the NLO level in QCD. Unstable resonant particles like the  $W^\pm/Z$ vector bosons and top quarks are modelled with Breit-Wigner propagators. Furthermore double-, single- and non-resonant contributions as well as all their interference terms are consistently included at the matrix element level. We   provide integrated and differential (fiducial) cross-section results and investigate the size of theoretical uncertainties as well as the impact of renormalization- and factorization scale settings on the  theoretical predictions.  Finally, we explore the effects of  additional jet activity  in $t\bar{t}W^+$ production and top-quark decays.}

\FullConference{42nd International Conference on High Energy Physics (ICHEP2024)\\
18-24 July 2024\\
Prague, Czech Republic\\}

\begin{document}
\maketitle

\def\WWW{e^+\nu_e\, \mu^-\bar{\nu}_\mu\, \tau^+\nu_\tau\, b\bar{b}}
\def\ub{\bar{u}}
\def\db{\bar{d}}
\def\sb{\bar{s}}
\def\cb{\bar{c}}
\def\tb{\bar{t}}

\section{Introduction}

Gluon initiated processes typically dominate $t\tb$ associated productions at the LHC. However, for $t\tb
W^\pm$ the $gg$ channel only opens up at higher orders leading to large real-emission
corrections. With approximate NNLO predictions only becoming available recently \cite{Buonocore:2023ljm},
alternative approaches have been developed to include such contributions.  For example, the
$t\tb W^\pm j$ process at NLO in QCD can be used to estimate their impact. In addition, the FxFx matrix
element merging procedure for $pp\to t\bar{t}W^\pm$ production at NLO in QCD with one and/or two additional
jets \cite{Frederix:2021agh} provide improved inclusive predictions. Top quark decays are not included in
the NNLO result, but are approximated using a parton-shower  programs in the FxFx prediction. 
Top and $W$ decays can be realistically modelled using full off-shell calculations. Such theoretical predictions have been previously calculated for the $pp \to t\tb W^\pm$
process \cite{Bevilacqua:2020pzy,Denner:2020hgg,Denner:2021hqi,Bevilacqua:2021tzp}. In this work we briefly summarize our
results where this treatment has been extended to include an additional jet \cite{Bi:2023ucp}.  In
the long-term these off-shell predictions  for $t\bar{t}W^\pm$ and $t\bar{t}W^\pm j$ can be
merged to obtain an improved description of the inclusive $t\tb W$ process in the multilepton
final state. In the following we shall shortly describe the state-of-the-art off-shell predictions for the
$t\tb Wj$ process, study the impact of NLO QCD corrections and estimate the size of various theoretical
uncertainties. Secondly, we will examine the impact of  additional jet activity  in  the $t\tb W$
process on the integrated and differential (fiducial) cross-section results. 

\section{Computational setup}

We consider the fully leptonic decay channel of the $pp \to t\bar{t}W^+j+X$ process at  ${\cal O} (\alpha_s^3 \alpha^6)$ and ${\cal O} (\alpha_s^4 \alpha^6)$ for the LHC with a center-of-mass energy of  $\sqrt{s}=13\,\textrm{TeV}$. Thus, we calculate NLO QCD corrections to the following LO final state $e^+ \nu_e \, \mu^-\bar{\nu}_{\mu} \, \tau^+ \nu_\tau\,  b \bar{b} j$. The three different lepton families are employed to simplify the calculation by avoiding $Z/\gamma\to \ell^\pm \ell^\mp$ contributions. The full off-shell approach includes all double-, single- and non-resonant contributions as well as their interference terms. Unstable resonant particles are described by the Breit-Wigner propagators and are treated using the complex mass scheme. The calculation is performed in the \textsc{Helac-Nlo} framework \cite{Bevilacqua:2011xh}, which employs  the  Dyson-Schwinger recursion  algorithm to efficiently calculate helicity amplitudes. The framework is split into two parts. First, \textsc{Helac-1Loop} \cite{vanHameren:2009dr} calculates the finite part of 1-loop amplitudes using the OPP reduction method \cite{Ossola:2006us} implemented in \textsc{CutTools} \cite{Ossola:2007ax}. The necessary scalar integral are provided by \textsc{OneLOop} \cite{vanHameren:2010cp}.  Secondly, \textsc{Helac-Dipoles} \cite{Czakon:2009ss,Bevilacqua:2013iha}, which implements two different subtraction schemes, is used to calculate real-emission contributions. The obtained results are stored in modified Les-Houches event files \cite{Alwall:2006yp} and \textsc{Root Ntuple} files \cite{Antcheva:2009zz}, which allows us to reweight events to different scale choices and  PDF sets without rerunning the expensive calculation. Predictions are provided for two dynamical  scale settings: $\mu_0=H_T/2$ and $\mu_0=E_T/2$, where $H_T$ and $E_T$ are defined as
\begin{equation}
H_T = p_T(e^+ )+ p_T(\tau^+)+ p_T(\mu^-) +p_{T}^{miss} + p_{T} (b_1) + p_{T} (b_2) + p_T(j_1) \,,
\label{eq:scale_ht}
\end{equation}
\begin{equation}
E_T = \sqrt{m_t^2 + p_{T}^2(t)} + \sqrt{m_t^2 + p_{T}^2 (\bar{t}\,)} + \sqrt{m_W^2 + p_{T}^2 (W)} 
+ p_T(j_1) \,,
\end{equation}
as well as for one fixed scale choice:  $\mu_0=m_t+m_W/2$. Following the recommendation from the PDF4LHC working group \cite{PDF4LHCWorkingGroup:2022cjn} we provide results for three modern PDF  sets using the \textsc{LHAPDF6} library \cite{Buckley:2014ana}. Specifically, we employ NNPDF3.1 \cite{NNPDF:2017mvq}, MSHT20 \cite{Bailey:2020ooq} and CT18 \cite{Hou:2019efy}. Scale uncertainties are  estimated  using  a 7-point scale variation. The internal PDF uncertainties are calculated using the prescriptions provided by the authors of the respective PDF set.

\section{Phenomenological results for $\boldsymbol{t\bar{t}W^+ j}$}

\begin{table}[t!]
    \centering
    \begin{tabular}{c | c c c c c c c}
    \midrule
        PDF & $\mu_0$ & $\sigma^{LO}\,$[ab] & $\delta_{scale}$ & $\sigma^{NLO}\,$[ab] & $\delta_{scale}$ & $\delta_{PDF}$ & $\mathcal{K}$\\
    \midrule
       NNPDF3.1 & $H_T/2$ & 115.8 & $^{+38\%}_{-26\%}$ & 142.3 & $^{+1.4\%}_{-8.1\%}$ & $^{+1.2\%}_{-1.2\%}$ & 1.23 \\
        & $E_T/2$ & 103.8 & $^{ +37 \% }_{ -25 \% }$ & 139.7 & $^{+3.7\%}_{-9.9\%}$ & $^{+1.2\%}_{-1.2\%}$ & 1.35\\
        & $m_t+m_W/2$ & 141.0 & $^{ +41 \% }_{ -27 \% }$ & 144.3 & $^{~+0.3\%}_{-14.1\%}$ & $^{+1.2\%}_{-1.2\%}$ & 1.02\\
    \midrule
    \end{tabular}
    \caption{ LO and NLO integrated (fiducial) cross-section results for the $pp\rightarrow \WWW j$ process calculated for the NNPDF3.1 PDF set and three different scale settings. Also given are the corresponding theoretical uncertainties and ${\cal K}$-factors.}
    \label{tab:fiducial_xs_nnpdf}
\end{table}

\begin{figure}[t!]
    \centering
    \includegraphics[width=0.45\linewidth]{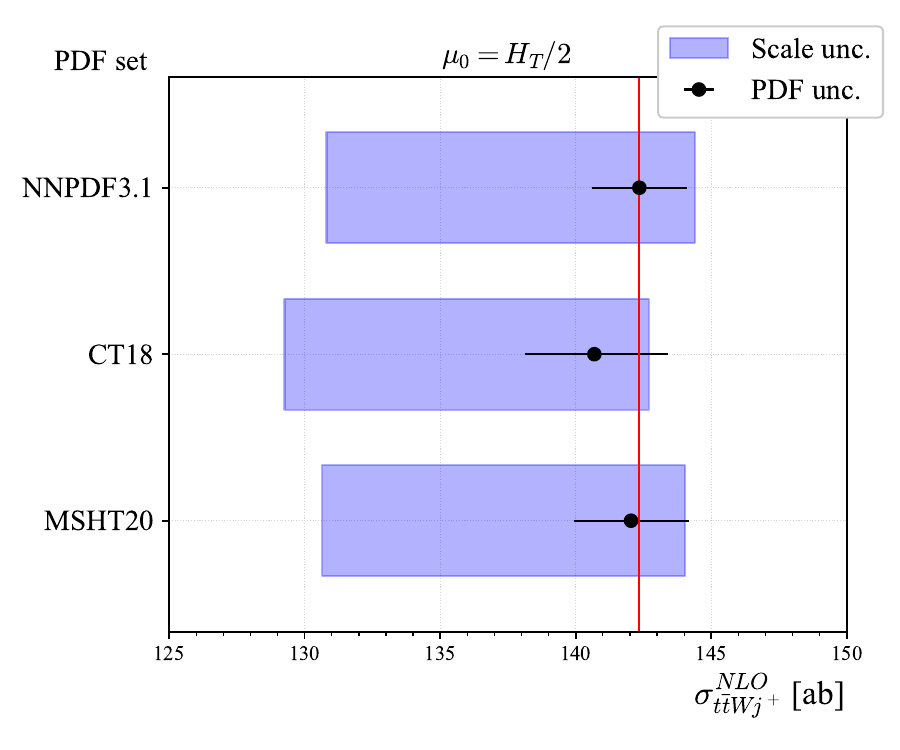}
    \includegraphics[width=0.45\linewidth]{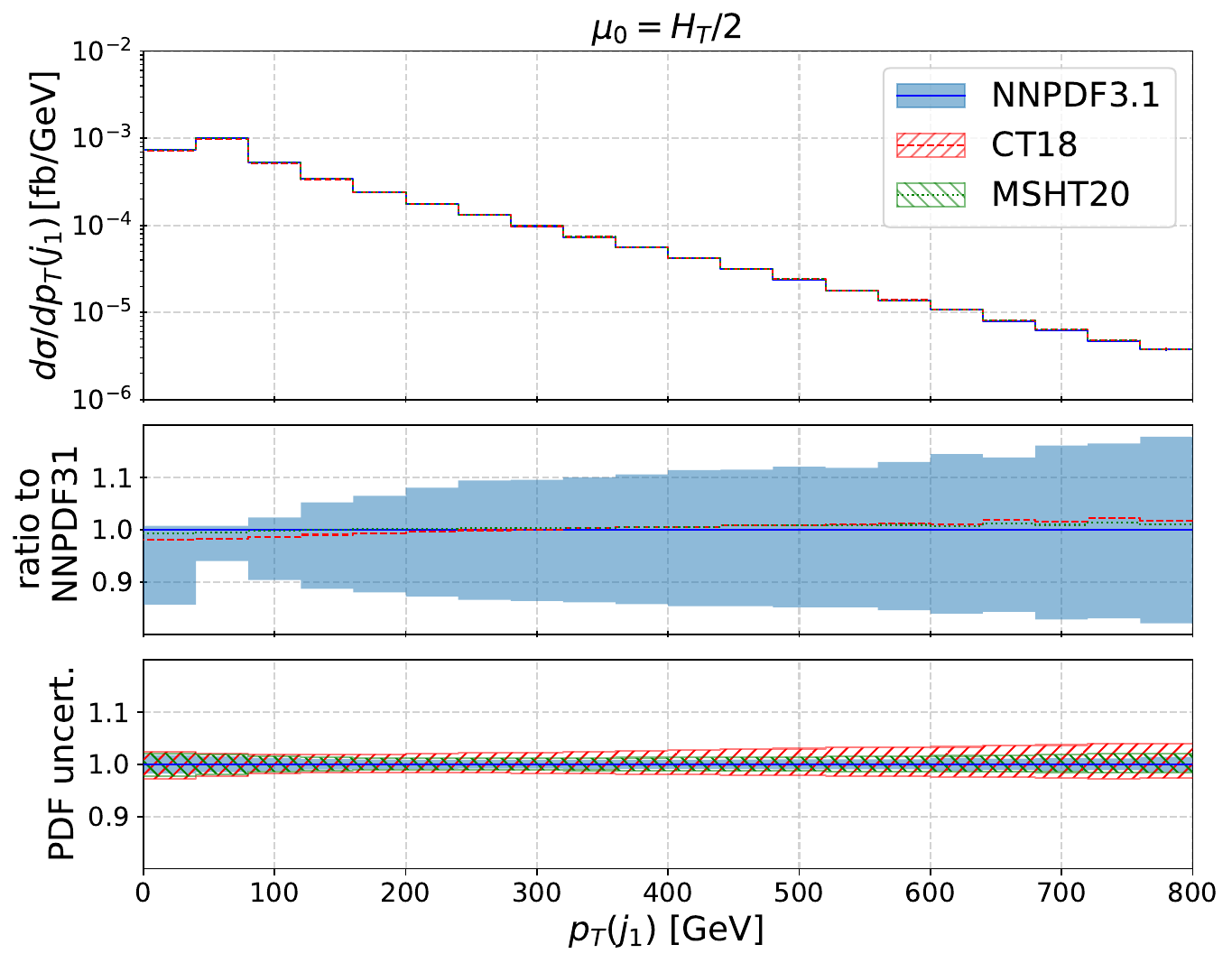}
    \caption{\it{Left: NLO integrated (fiducial) cross-section results for the $pp\rightarrow \WWW j$ process with the theoretical uncertainties from scale dependence and PDFs, obtained for the  NNPDF3.1, CT18 and MSHT20 PDF sets using $\mu_0=H_T/2$. Right: Differential distribution for $p_T(j_1)$ for the same three PDF sets  and $\mu_0$ together with their corresponding theoretical uncertainties. }}
    \label{fig:xs_uncert_comparison}
\end{figure}

We start with the LO and NLO fiducial cross-section results  for the $pp\rightarrow \WWW j$ 
process for our three different scale choices. They are presented in Table \ref{tab:fiducial_xs_nnpdf}. Also given there are the corresponding theoretical uncertainties and ${\cal K}$-factors. NLO corrections  are in the range of $2\%-35\%$ and are covered by the large $(40\%)$ uncertainties associated with the LO predictions. At NLO  theoretical uncertainties are reduced to $8\%-10\%$  for the dynamical scale settings,  and to $14\%$ for the fixed scale choice. The spread in the central value of the predictions at the LO is large, but still within the substantial LO uncertainties. At NLO the predictions stabilize and the differences among the various scale choices reduce to about $3\%$. Internal PDF uncertainties are negligible at only about $1\%$. A graphical representation of these findings  is provided  in Figure \ref{fig:xs_uncert_comparison}. Furthermore, we find that the results obtained with the three PDF sets are compatible within their internal uncertainties. 
\begin{figure}
    \centering
    \includegraphics[width=0.45\linewidth]{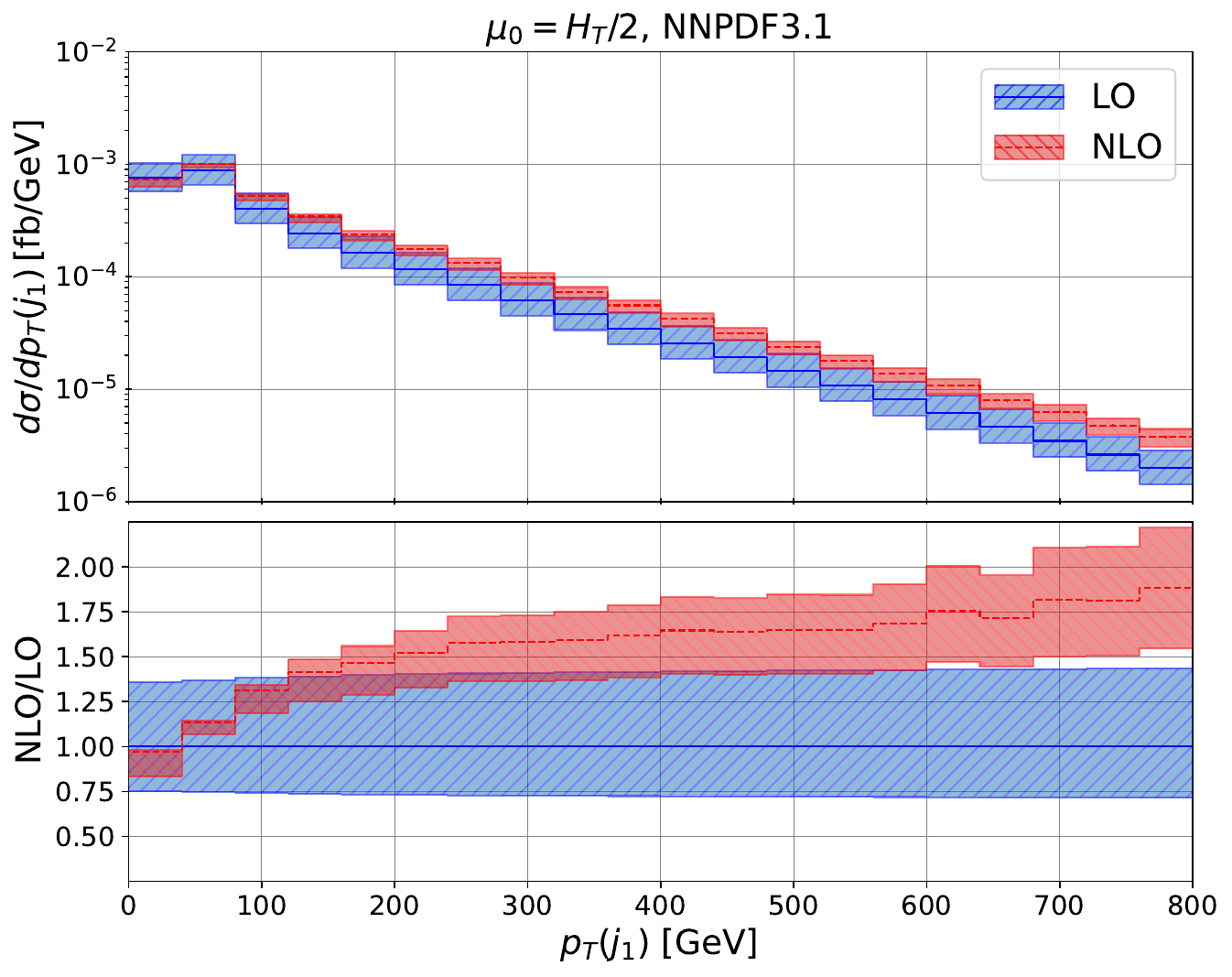}
    \includegraphics[width=0.45\linewidth]{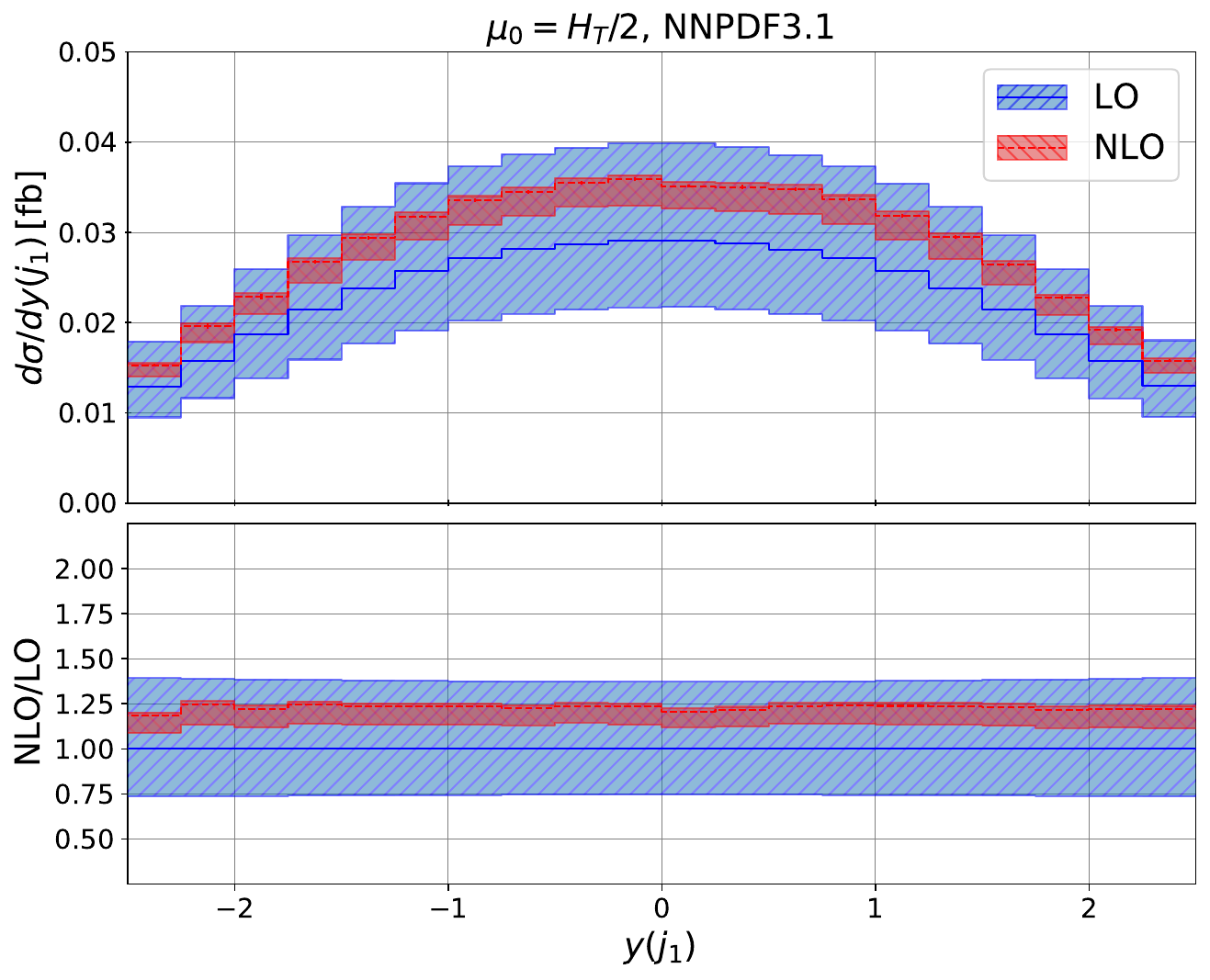}
    \caption{LO and NLO differential cross-section distribution as a function of  the transverse momentum of the hardest light jet, $p_T(j_1)$,  and rapidity of the hardest light jet, $y(j_1)$, for the  $pp \to \WWW \, j+X$ process. Also given are the theoretical uncertainties coming from the scale dependence. Lower panels display the differential ${\cal K}$-factors together with their uncertainties.}
    \label{fig:kfactor}
\end{figure}

On the differential level NLO corrections can be large in specific phase-space regions. An example
is presented in Figure \ref{fig:kfactor}. We can observe that the transverse momentum of the hardest light jet, $p_T(j_1)$, exhibits  higher-order QCD corrections up to $80\%$ towards the tail of the distribution. For dimensionless observables, like the rapidity of the hardest light jet, $y(j_1)$, we find instead almost flat NLO corrections. Including higher-order corrections reduces the scale uncertainties to about $18\%$ and $10\%$ for $p_T(j_1)$ and $y(j_1)$ respectively. Also at the differential cross-section level the PDF uncertainties remain negligible when compared to the theoretical uncertainties from the scale dependence.

\section{Additional jet activity in $\boldsymbol{t\bar{t}W^+}$}

\begin{figure}
    \centering
    \includegraphics[width=0.45\linewidth]{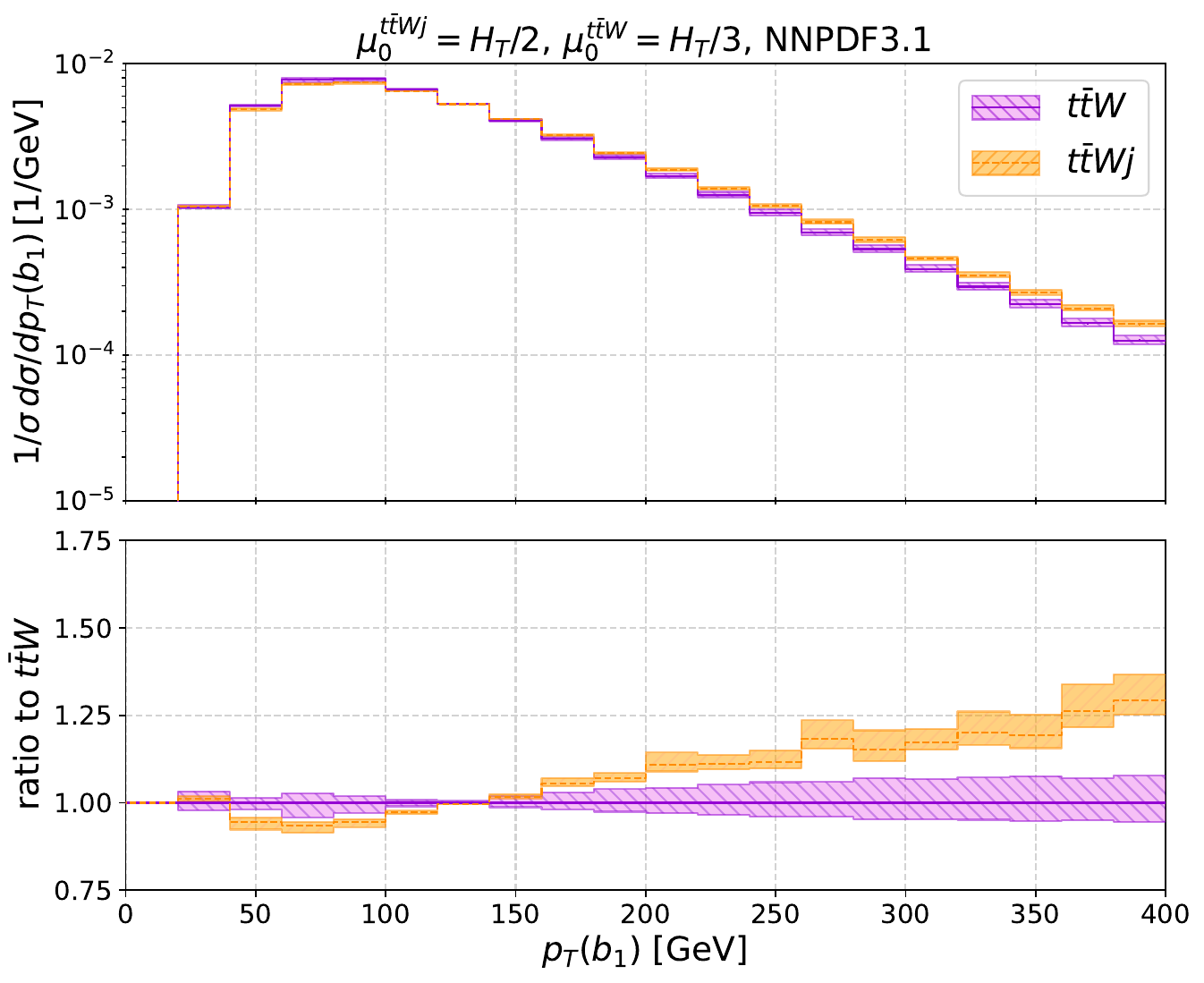}
    \includegraphics[width=0.45\linewidth]{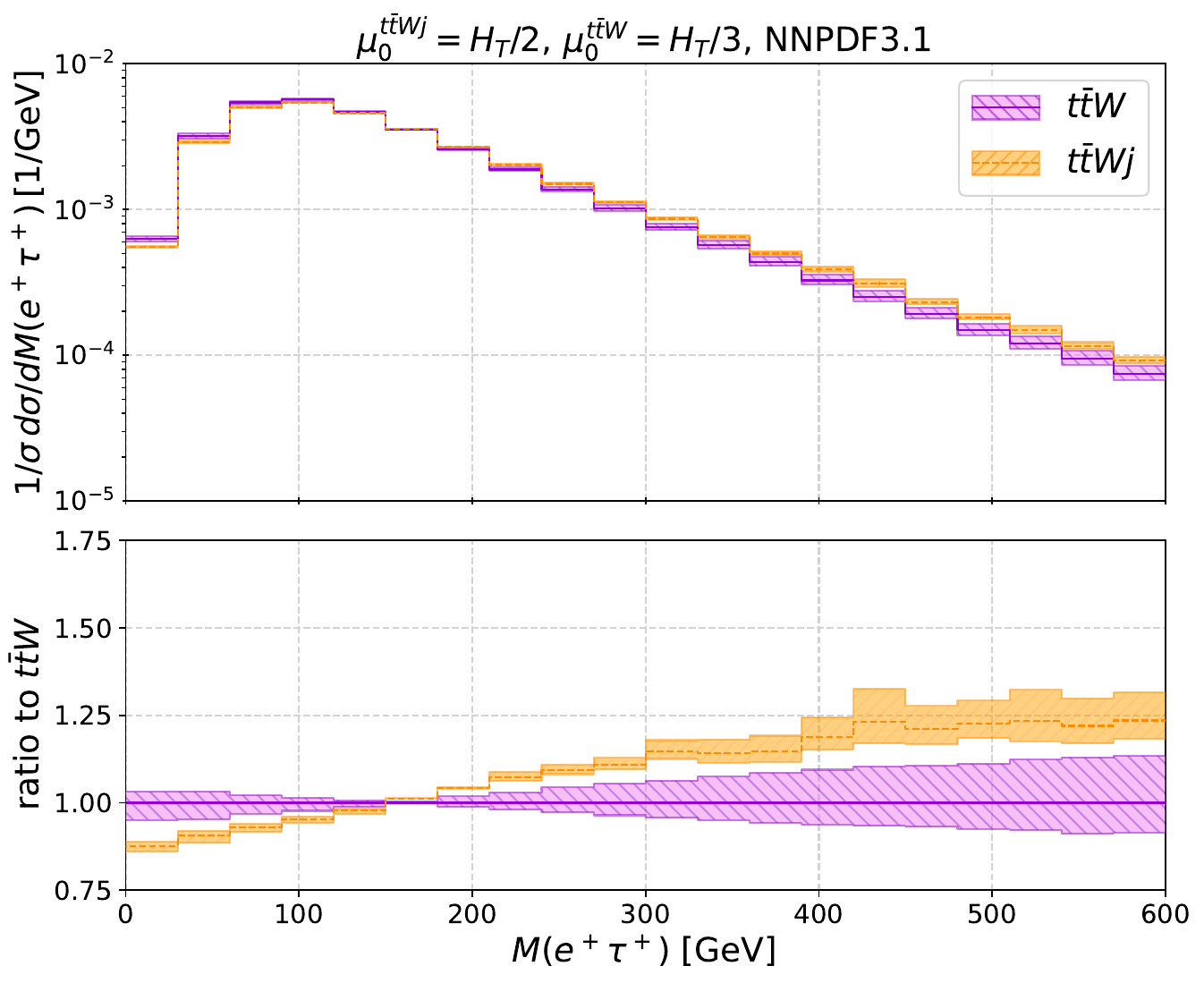}
    \caption{NLO normalized differential cross-section distributions as a function of the transverse momentum of the hardest $b$-jet, $p_T(b_1)$, and the invariant mass of the same-sign dilepton pair, $M_{e^+\tau^+}$, for  the $pp \to \WWW \, +X$   and $pp \to \WWW \, j+X$$ $  processes. Also given are the theoretical uncertainties coming from the scale dependence. Lower panels display the ratios to the $pp \to \WWW \, +X$  result. }
    \label{fig:ttw_ttwj}
\end{figure}

The LO and NLO integrated (fiducial) cross-section results for the $pp \to \WWW \, +X$  and the $pp \to \WWW \, j+X$  processes are given in Table \ref{tab:ttwttwjfiducial}. Depending on the exact scale choice, $pp \to \WWW \, j+X$ represents about $50\%$  of the contribution to the overall $pp \to \WWW \, +X$  process.  This underscores the need to take into account additional jet activity already at the fiducial cross-section level by means of merging techniques or NNLO calculations. In order to assess the impact also on the  differential cross-section level  we compare the shapes of  the following two  observables: the transverse momentum of the hardest $b$-jet, $p_T(b_1)$, and the invariant mass of the same-sign dilepton pair, $M_{e^+\tau^+}$. Both NLO (normalised) differential cross-section distribution  are displayed  in Figure \ref{fig:ttw_ttwj}. We find that the shapes of the distributions are altered throughout the entire plotted range. For $p_T(b_1)$ differences of up to $30\%$ can be noticed and for $M_{e^+\tau^+}$ they are rather similar up to about $25\%$. In both cases, they are well outside of the respective uncertainty bands.
In summary, additional light jets are required to accurately describe the shape of these quantities for the  $pp \to \WWW \, +X$ process. For dimensionless observables, we find that these effects are usually much less pronounced. 
\begin{table}
    \centering
    \begin{tabular}{l l l l l l l}
        \hline\hline
        &&&&&&\\[-0.2cm]
         & $\sigma^{t\bar{t} W^+}_{H_T/3}\,[\textrm{ab}]$ & $\sigma^{t\tb W^+j}_{H_T/2}\,[\textrm{ab}]$ & $\sigma^{t\tb W^+}_{E_T/3}\,[\textrm{ab}]$ & $\sigma^{t\tb W^+j}_{E_T/2}\,[\textrm{ab}]$ & $\sigma^{t\tb W^+}_{m_t+m_W/2}\,[\textrm{ab}]$ & $\sigma^{t\tb W^+j}_{m_t+m_W/2}\,[\textrm{ab}]$\\
         &&&&&&\\[-0.2cm]
         \hline\hline
         &&&&&&\\[-0.2cm]
     LO  & $ 216.6^{\,+24\%}_{\,-18\%} $ 
     & $ 115.8^{\,+38\%}_{\,-26\%} $ 
     & $ 198.7^{\, +23 \% }_{ \,-18 \% }$ 
     & $103.9^{\, +37 \% }_{ \,-25 \% }$ 
     & $ 202.6^{ \,+24 \% }_{\, -18 \% }$ 
     & $141.0^{ \,+41 \% }_{ \,-27 \% }$\\[0.2cm]
     NLO & $254.6^{ \,+2.8 \% }_{ \,-5.9 \% }$ 
     & $ 142.3^{\, +1.4 \% }_{ \,-8.1 \% }$ 
     & $ 249.6^{ \,+4.6 \% }_{\, -6.8 \% }$ 
     & $ 139.7^{ \,+3.7 \% }_{\, -9.9 \% }$ 
     & $ 252.3^{\, +4.5 \% }_{ \,-6.8 \% }$ 
     & $144.3^{\, +~0.3 \% }_{ \,-14.1 \% }$\\
     &&&&&&\\[-0.2cm]
     \hline\hline
    \end{tabular}
    \caption{\it LO and NLO integrated (fiducial) cross-section results for the $pp \to \WWW \, +X$   and $pp \to \WWW \, j+X$$ $ processes calculated for the NNPDF3.1 PDF set and three different scale settings. Also given are the corresponding scale uncertenties.}
    \label{tab:ttwttwjfiducial}
\end{table}

\section{Conclusion}

We calculated the NLO QCD predictions for the $pp \to \WWW \, j+X$ process with full off-shell effects included. We found that at the integrated (fiducial) cross-section level NLO corrections are small to moderate, depending on the LO PDF set used. At the differential cross-section level, they can even reach $80\%$ in certain the phase-space regions.  The PDF uncertainties are negligible when comparing to the scale uncertainties. Additional jet activity in the $pp \to \WWW +X$ process leads to large changes in the shape of the differential cross-section distributions. This indicates that these  jets must be included in the description of this process. Therefore, we plan to combine our full off-shell predictions for $pp \to \WWW +X$ and $pp \to \WWW \, j+X$  to improve the description of the $pp \to \WWW +X$ process at NLO in QCD.

\FloatBarrier

\bibliographystyle{JHEP}

\providecommand{\href}[2]{#2}\begingroup\raggedright\endgroup

\end{document}